\makeatletter
\def\input@path{{./tex/}{./bst/}}
\makeatother
\documentclass[%
reprint,
superscriptaddress,
 amsmath,amssymb,
 aps,
 prx,
floatfix,
]{revtex4-2}

\usepackage{graphicx}
\usepackage{dcolumn}
\usepackage{bm}
\usepackage{hyperref}
\usepackage{siunitx}
\usepackage{physics}
\usepackage{upgreek}
\usepackage{comment}
\usepackage{subscript}
\usepackage{xprintlen}
\usepackage{ulem}

\usepackage[dvipsnames]{xcolor}  

\newcommand{\mose}{MoSe$_2$}

\newcommand{\Xbot}{X$_\text{bot}$}

\newcommand{\RPbot}{RP$_\text{bot}$}

\newcommand{\ct}[1]{\textcolor{green}{\textbf{[cite!]}}}

\newcommand{\abbrev}[3]{
  \newcounter{#3}
  \setcounter{#3}{0}
  \newcommand{#1}{\ifnum\value{#3}<1{#2 (#3)}\else{#3}\fi\stepcounter{#3}}}

\abbrev{\hBN}{hexagonal boron nitride}{h-BN}
\abbrev{\TMD}{transition metal dichalcogenide}{TMD}
\abbrev{\AP}{attractive polaron}{AP}
\abbrev{\RP}{repulsive polaron}{RP}
\abbrev{\DR}{differential reflection}{DR}
\abbrev{\RF}{resonance fluorescence}{RF}
\abbrev{\vdW}{van der Waals}{vdW}
\abbrev{\LED}{light-emitting diode}{LED}
\abbrev{\CW}{continuous wave}{CW}
\abbrev{\CCD}{charge-coupled device}{CCD}
\abbrev{\APD}{avalanche photodiode}{APD}

\begin{document}


\title{Photon Correlation Spectroscopy as a Probe of Critical Fluctuations in Correlated Electron Systems}

\author{Natasha Kiper}
\affiliation{Institute for Quantum Electronics, ETH Zürich, Zürich, Switzerland}
\author{Bertrand Evrard}
\affiliation{Institute for Quantum Electronics, ETH Zürich, Zürich, Switzerland}
\affiliation{Sorbonne Université, CNRS, Institut des Nanosciences de Paris, Paris, France}
\author{Yuya Shimazaki}
\affiliation{Institute for Quantum Electronics, ETH Zürich, Zürich, Switzerland}
\affiliation{Institute for Solid State Physics, The University of Tokyo, Kashiwa, Chiba 277-8581, Japan}
\author{Ido Schwartz}
\affiliation{Institute for Quantum Electronics, ETH Zürich, Zürich, Switzerland}
\affiliation{The Physics Department and the Solid State Institute, Technion – Israel Institute of Technology, 32000 Haifa, Israel}
\author{Martin~Kroner}%
\affiliation{Institute for Quantum Electronics, ETH Zürich, Zürich, Switzerland}
\author{Kenji Watanabe}
\affiliation{Research Center for Electronic and Optical Materials, National Institute for Materials Science, 1-1 Namiki, Tsukuba 305-0044, Japan}
\author{Takashi Taniguchi}
\affiliation{Research Center for Materials Nanoarchitectonics, National Institute for Materials Science,  1-1 Namiki, Tsukuba 305-0044, Japan}
\author{Atac {\.I}mamo{\u{g}}lu}
\affiliation{Institute for Quantum Electronics, ETH Zürich, Zürich, Switzerland}

\date{\today}

\begin{abstract}
Characterizing phase transitions between correlated electronic phases, extracting their critical exponents, and identifying their universality class are of central interest in many-body physics. Here, we propose and demonstrate that photon correlation spectroscopy can be used to gain insight into the nature of critical electronic density fluctuations. We study a semiconductor moiré material consisting of two \mose\ layers separated by a monolayer \hBN\ spacer and measure interlayer electron dynamics via the second-order correlation function of the scattered photons. The correlated transfer of large numbers of electrons between the layers at the onset of an Ising-type layer pseudo-spin phase transition leads to photon bunching in light scattered by the exciton resonance of one layer. Our measurements pave the way for using photon correlations as a method to access dynamical exponents associated with electronic phase transitions.
\end{abstract}
\maketitle

\section{Introduction} \label{sec:introduction}

Photon correlation measurements have played a central role in the development of quantum optics. While the measurements of Hanbury-Brown and Twiss (HBT) demonstrated the presence of excess fluctuations associated with the bosonic nature of photons~\cite{brownCorrelationPhotonsTwo1956}, observation of photon antibunching in resonance fluorescence by Kimble, Dagenais and Mandel~\cite{kimblePhotonAntibunchingResonance1977} unequivocally established the quantum nature of light fields generated by single quantum emitters. Beyond providing conceptual advances, photon correlations currently play a central role in applications ranging from quantum key distribution to linear optics quantum computation. Even though the focus of scientific research and the aforementioned applications has been to use photon correlations to assess the {\sl quantumness} of the detected light field, it is equally valid to regard photon antibunching as providing a statement about the quantumness of the material that is used to generate the photons~\cite{nambiarDiagnosingElectronicPhases2025,kassManyBodyPhotonBlockade2024}. In this perspective, photon antibunching in emission from a single atom can be understood as stemming from the suppressed fluctuations in atomic polarization due to the anharmonic nature of its spectrum. When the number of atoms or electrons that make up the light source fluctuates in time~\cite{hennrichTransitionAntibunchingBunching2005,ozturkFluctuationDissipationRelationBoseEinstein2023}, or when the electronic polarization has a thermal distribution~\cite{morganMeasurementPhotonBunching1966}, the generated field generically exhibits excess fluctuations, which leads to photon bunching.

Motivated by this second perspective, we establish photon correlation measurements as an experimental condensed-matter technique~\cite{nambiarDiagnosingElectronicPhases2025} to investigate fluctuations in electron density across a phase transition. In particular, the excess fluctuations at the transition temperature and their characteristic dynamic power-law exponent in a classical second-order phase transition could be directly measured by monitoring the decay of the bunching peak as a function of the delay time between the successive photon detection events. While we focus on a classical phase transition, the technique we develop could also be applied to a quantum phase transition, where the back-action from light scattering would induce fluctuations in the monitored electronic observable, and would in turn constitute the source of electron or photon number fluctuations at the quantum critical point even at vanishing electronic temperature. We also emphasize that photon correlation measurements were previously used to study a wide range of problems not only in photonics~\cite{finkSignaturesDissipativePhase2018}, but more prominently in chemistry~\cite{foordDeterminationDiffusionCoefficients1970} and material science~\cite{sidebottomConnectingStructureDynamics2007, madsenSimpleExponentialCorrelation2010, sinhaXrayPhotonCorrelation2014, madsenStructuralDynamicsMaterials2016}.

In recent years, van der Waals heterostructures incorporating transition metal dichalcogenides, with and without moiré potentials, have emerged as a fertile platform to study electronic phases with strong correlations, such as Mott states~\cite{reganMottGeneralizedWigner2020, tangSimulationHubbardModel2020,shimazakiStronglyCorrelatedElectrons2020,wangCorrelatedElectronicPhases2020}, Wigner~\cite{smolenskiSignaturesWignerCrystal2021,zhouBilayerWignerCrystals2021} or generalized Wigner crystals~\cite{reganMottGeneralizedWigner2020,xuCorrelatedInsulatingStates2020}, the (fractional) quantum anomalous Hall effect~\cite{liQuantumAnomalousHall2021, caiSignaturesFractionalQuantum2023, zengThermodynamicEvidenceFractional2023,xuObservationIntegerFractional2023,parkObservationFractionallyQuantized2023}, kinetic magnetism~\cite{ciorciaroKineticMagnetismTriangular2023}, and superconductivity~\cite{xiaSuperconductivityTwistedBilayer2025, guoSuperconductivity50degTwisted2025, xiaBandwidthtunedMottTransition2026}. The nature of phase transitions between correlated and trivial, as well as between competing correlated phases, is a topic of ongoing research~\cite{tangDielectricCatastropheWignerMott2022, zhouQuantumMeltingGeneralized2024,tsuiDirectObservationMagneticfieldinduced2024, xiangImagingQuantumMelting2025, liuFractionalChernInsulators2025}. Characterizing these phase transitions, extracting their critical exponents and identifying their universality class is a recurring and often outstanding challenge of many-body physics. In particular, the theory of quantum criticality assigns a crucial role to the dynamical exponent $z$: Quantum  phase transitions can be mapped onto classical ones, with an extra ``imaginary time" dimension, scaling as $\tau\sim\xi^{z}$ with $\xi$ the spatial correlation length \cite{sachdevQuantumPhaseTransitions1999,vojtaQuantumPhaseTransitions2003}. Direct experimental access to the critical dynamics is therefore essential to bridge quantum and classical critical behaviors.

Conventional optical and transport measurements in \TMD\ heterostructures probe the steady state of the system. To probe dynamics, it is necessary to use time-resolved methods. One such method that has been extensively used to study \vdW\ heterostructures with \TMD s is pump-probe spectroscopy~\cite{conteUltrafastPhotophysics2D2020}, in which a high-intensity pump pulse perturbs the system, which is then probed by a second, low-energy probe pulse. However, this method relies on bringing the system out of equilibrium using the strong pump pulse. Here, we present a different time-resolved method that can be used to measure dynamics of the electronic state itself in thermal equilibrium, without relying on inducing excitations using an optical pump. We establish our approach by revealing the critical fluctuations of a bilayer Mott insulator in the vicinity of the critical temperature at which the electrons become layer-polarized.

\section{Photon correlation measurements as a probe of electron density fluctuations}

\subsection{Description of photon correlation spectroscopy}

Photon correlation spectroscopy relies on measuring the second-order autocorrelation function $g^{(2)}(\tau)$ of the photons scattered by the sample. It is defined as
\begin{equation}
	g^{(2)}(\tau)=\frac{\langle : I(t)I(t+\tau) :\rangle}{\langle I(t)\rangle^2},
\end{equation}
where $I(t) = \langle \hat{a}_\text{out}^\dagger\hat{a}_\text{out} \rangle$ is the intensity of the detected light beam at time $t$, $\hat{a}_\text{out}$ denotes the annihilation operator of the detected photon mode, the angular brackets signify the time or ensemble average and $:\;:$ denote normal ordering of the operators. Since we are interested in classical light sources driving an emitter composed of a large number of electrons, the normal ordering of the operators is not relevant. The value of $g^{(2)}(\tau)$ measures  conditional probability of detecting a photon at time $t+\tau$, given that another photon was detected at time $t$. At sufficiently long times, the correlations must vanish and therefore $g^{(2)}(\tau\to\infty)=1$. In coherent light beams, such as those generated by lasers, the photons are completely uncorrelated; $g^{(2)}(\tau)=1$ $\forall \tau$. If $g^{(2)}(0) > 1$, a detection event increases the probability of a second photon detection. This is referred to as photon bunching.

\begin{figure}[h]
    \includegraphics[width=\columnwidth]{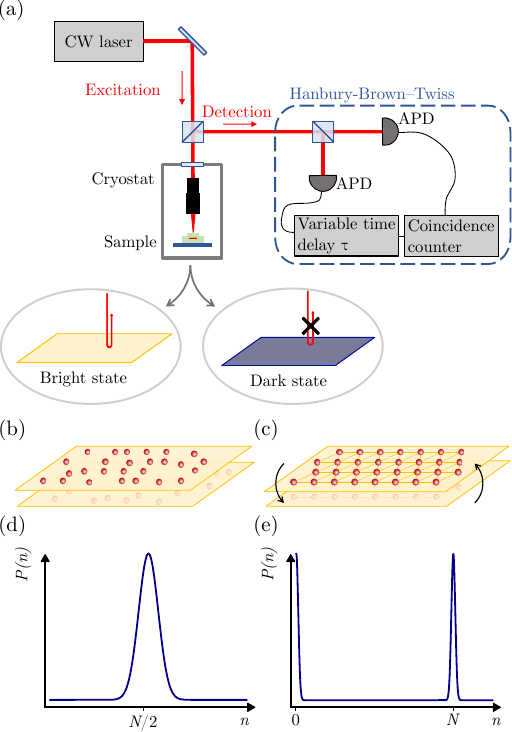}
    \caption{Schematic illustration of the experiment. (a) Sketch of the measurement setup. Light emitted from a CW laser is focused on the sample inside a cryostat. Depending on the electronic state of the TMD a larger or smaller fraction of the excitation light is reflected. This reflected light is analyzed using a Hanbury-Brown--Twiss setup consisting of a beam splitter, two APDs, and a coincidence counter, allowing the extraction of the second-order correlation function of reflected photons. (b) Illustration of an uncorrelated state of electrons in a bilayer TMD system. (c) Illustration of a bilayer TMD system hosting a layer-polarized Mott state. (d-e) Corresponding probability densities $P(n)$ of finding an electron density $n$ within the optical spot in one layer of the bilayer system. For a Fermi liquid state (panels b, d), the particles are uncorrelated and thermal fluctuations lead to a Gaussian electron number distribution. If electrons form a layer-polarized Mott--Wigner state (panels c, e), they exhibit correlated motion between the layers when the energy levels in the two layers are degenerate. This leads to a bimodal distribution and enhanced density fluctuations.
    }
    \label{fig:1}
\end{figure}

The general principle of photon correlation spectroscopy on a \TMD\ system is illustrated in Fig.~\ref{fig:1}. The core idea is to detect electron density fluctuations in a \TMD\ layer which lead to fluctuations of the light scattered from that layer. Specifically, light scattered from the \AP\ or \RP\ resonances carries information about the instantaneous charge density in the layer hosting the resonance. The polaron resonances can be described by the Chevy Ansatz~\cite{chevyUniversalPhaseDiagram2006}
\begin{align}
    \ket{\Psi_0}=(\phi_0\hat{x}^\dagger+\sum_\textbf{k,q}\psi_\textbf{k,q}\hat{x}^\dagger_\textbf{q-k}\hat{e}^\dagger_\textbf{k}\hat{e}_\textbf{q})\ket{\text{FS}},
\end{align}
where $\hat{x}_\textbf{q}^\dagger$ and $\hat{e}_\textbf{q}^\dagger$ ($\hat{x}_\textbf{q}$ and $\hat{e}_\textbf{q}$) are the creation (annihilation) operators for an exciton and electron at momentum \textbf{q}, respectively, and $\ket{\text{FS}}$ denotes the wavefunction of the unperturbed Fermi sea. We have assumed electron doping, but the physics is analogous for hole doping. 

Equivalently, the polaron creation operator $\hat{p}^\dagger_{i}$ with $i\in\{\text{AP},\text{RP}\}$  can be written as
\begin{align}
    \hat{p}^\dagger_{i}=\phi_0^i \hat{x}^\dagger+\sum_\textbf{k,q}\psi_{\textbf{k,q}}^i \hat{x}^\dagger_{\textbf{q-k}}\hat{e}^\dagger_\textbf{k}\hat{e}_\textbf{q}.
\end{align}
The oscillator strength, and therefore the coupling to the radiation field, is given by the polaron's exciton content parametrized by the quasiparticle weight $\phi_0^i$. 

Fluctuations in the electron density $n_e$ affect the light scattering rate through fluctuations of the quasiparticle weight $\phi_0^i$ or the polaron resonance energy. Although information on electron density fluctuations could be obtained through \AP\ or \RP, here we focus on the \RP\ in the low doping limit ($n_e\rightarrow 0$), where the \RP\ smoothly connects to the bare exciton, $\phi_0^{RP}\rightarrow1$. We therefore set $\hat{p}_\text{RP}\approx\hat{x}$. For the mode $\hat{a}_\text{out}$ of the light scattered at a detuning $\Delta \omega (n_e)$ from the resonance, we can write
\begin{align}
    \ev{\hat{a}_\text{out}^\dagger\hat{a}_\text{out}}= \frac{\chi\Gamma^2/4}{\Gamma^2/4+\Delta\omega(n_e)^2}\ev{\hat{x}^\dagger \hat{x}}.
\end{align}
Here, $\Gamma$ is the exciton decay rate and $\chi$ gives the efficiency with which an exciton is converted into a photon.
We linearize the effect of electron density fluctuations $\delta n_e$ on the resonance energy $\omega_0$:
\begin{align}
    \omega_0^\prime=\omega_0+\eta\cdot\delta n_e,
\end{align}
with $\eta$ a small proportionality factor. We therefore find
\begin{align}
    \delta\ev{\hat{a}_\text{out}^\dagger\hat{a}_\text{out}}(t) =\frac{\chi\Gamma^2/4}{\Gamma^2/4+\Delta\omega_0^2}\ev{\hat{x}^\dagger\hat{x}}\cdot\eta\cdot\delta n_e(t),
\end{align}
showing that photon number fluctuations of the scattered light are linearly proportional to the electron number fluctuations. Here, $\Delta \omega_0$ is the detuning of the probe laser from the bottom layer \RP\ resonance in the limit of vanishing $\delta n_e$. Since $\Gamma \sim 10^{12}$~s$^{-1}$ is much shorter than the electronic time scales of interest, the fluctuations in the detected intensity would adiabatically follow $\delta n_e(t)$.

Electron number fluctuations therefore directly map to fluctuations in scattered light intensity. An intuitive way to understand how this mapping leads to bunching in photon correlation measurements is to note that electron density fluctuations that result in a reduction of the instantaneous detuning $\Delta \omega (n_e(t))$ increase the photon detection probability at time $t$. For $\tau$ short compared to the characteristic timescales of electronic density fluctuations, the conditional probability of detecting a second photon at time $t+\tau$ is therefore enhanced by the first detection event at $t$. In the $g^{(2)}$ function, this enhancement manifests itself a bunching signal, $g^{(2)}(0)>1$. The amplitude and delay time dependence of the bunching signal are directly related to the amplitude and characteristic time scale of electron number fluctuations in the system, as our previous discussion shows.  Due to the spin-valley locking and optical selection rules in \TMD s, measuring in a circularly polarized basis would also allow for detecting the fluctuations in spin density.

\subsection{Experiments}

To demonstrate the photon correlation spectroscopy technique, we use the device presented in Refs.~\cite{shimazakiStronglyCorrelatedElectrons2020,shimazakiOpticalSignaturesPeriodic2021,schwartzElectricallyTunableFeshbach2021} as a model system. It contains two \mose\ monolayers with a relative twist angle of 0.8$^\circ$, separated by a monolayer \hBN\ spacer. A moiré pattern with lattice size $a_\text{m}\approx 24$~nm is formed by the two \mose\ layers and leads to a superlattice potential in both \TMD s. The \hBN\ spacer prevents coherent electron tunneling between the \TMD s and ensures that the \mose\ layers retain a direct band gap. Due to differential-strain-induced energy splitting of the exciton resonances, the optical response of \RP\ resonances of both layers can be measured independently.

The potential minima for electrons are at \textit{MX} and \textit{XM} sites, i.e., the high-symmetry stacking regions where metal (in this case, Mo) and chalcogen (in this case, Se) atoms are vertically aligned. The \textit{MX} and \textit{XM} sites together form a honeycomb lattice, but importantly, adjacent minima are located in different layers: electrons at \textit{MX} sites (Mo above Se$_2$) are located in the top layer, those at \textit{XM} sites (Se$_2$ above Mo) in the bottom layer. At a filling factor $\nu=1$, where $\nu$ denotes the number of electrons per moiré unit cell~\footnote{We note that we use different, but equivalent conventions for the values of $\nu$ and $V_E$ than Ref.~\cite{shimazakiStronglyCorrelatedElectrons2020}.}, on-site electronic repulsion leads to the formation of a Mott insulator~\cite{shimazakiStronglyCorrelatedElectrons2020}. There exist two equivalent layer-polarized Mott states with triangular charge arrangement, where all electrons either occupy bottom layer \textit{XM} sites or top layer \textit{MX} sites. A transition from one state to the other can be driven by the out-of-plane electric field applied by the gates. Due to the lateral displacement between \textit{MX} and \textit{XM} sites, Coulomb repulsion between nearest neighbor moiré sites favors layer-polarized states. Since the layer degree of freedom of electrons can be considered as a pseudo-spin, we can interpret the preference of the system to layer polarize as an effective Ising ferromagnetic interaction. This analogy suggests that the electric field applied by the top and bottom gates plays the role of a magnetic field.

At sufficiently low temperatures, the system should therefore undergo a continuous phase transition from a para- to a ferromagnetic state. However, we do not observe  hysteresis of the layer-polarization as we sweep the electric field, suggesting that the electronic temperature is still above the layer pseudospin Curie temperature. The sharpness of the electric-field-driven layer transition suggests that the electrons at $T=4$~K are in a critical regime, where we expect domains of layer-polarized electrons with a characteristic size $\xi\gg a_\text{m}$. Concurrently, we expect the domain dynamics to be much slower than the natural timescales of the electronic system. The conjunction of these two behaviors, together with the sensitivity of the optical resonances to the layer polarization, render this system ideal for showcasing the method of photon correlation spectroscopy.

\begin{figure}[h]
    \includegraphics[width=\columnwidth]{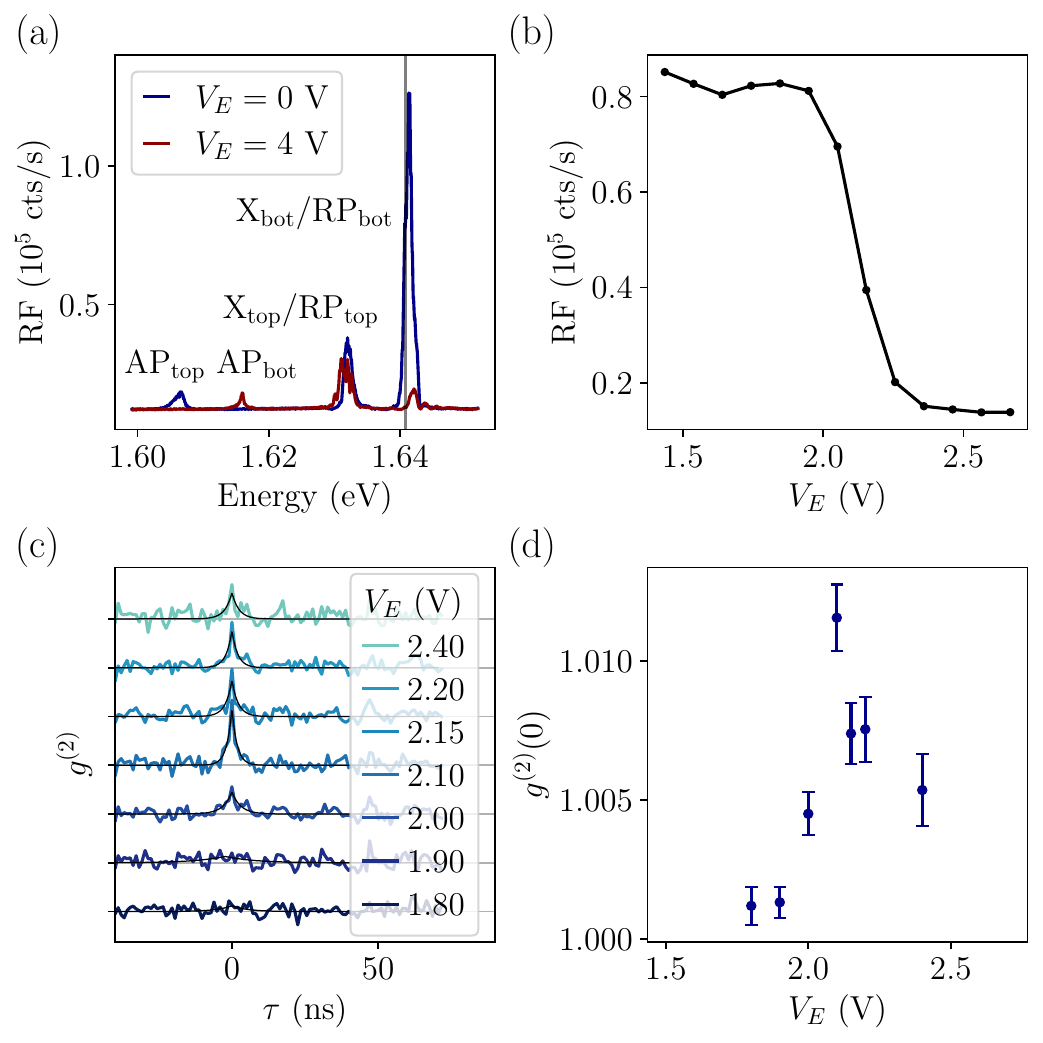}
    \caption{(a) Resonance fluorescence spectra taken in cross-linear polarization at total filling factor $\nu=1$ on both sides of the electric-field-driven layer transition. For $V_E=0$~V, the system is completely bottom-layer polarized (blue curve), while it is top-layer polarized for $V_E=4$~V (red curve). (b) Resonance fluorescence signal close to the \Xbot/\RPbot\ resonance (indicated by the gray vertical line in panel a) as a function of electric field $V_E$. At the layer transition close to $V_E=2.1$~V, the signal changes drastically. (c) Second-order correlation function as a function of time delay $\tau$ for different electric fields $V_E$. A bunching signal ($g^{(2)}(0)>1$) is clearly observed close to the phase transiton. The black lines are fits to the data. (d) Bunching amplitudes as a function of $V_E$, obtained by fitting the data shown in panel c. The bunching is maximal at the center of the phase transition, $V_E=2.1$~V.
    }
    \label{fig:2}
\end{figure}

We carry out photon correlation measurements in \RF\ configuration, where we excite and detect with orthogonal linearly polarized light. The total filling factor of the moiré lattice is $\nu = 1$, where the electrons form an incompressible Mott state~\cite{shimazakiStronglyCorrelatedElectrons2020}. Resonance fluorescence spectra on both sides of the layer transition at $\nu=1$ are shown in Fig.~\ref{fig:2}(a). These measurements are done without external magnetic field, where ideally, we would not expect any signal in cross-linear polarized detection. However, strain in the \TMD\ leads to a small but finite mixing of the valley degree of freedom of the excitons. The \RF\ signal from the bottom layer exciton in this sample is much stronger than that of the top layer exciton, indicating that the bottom layer has larger strain.

We use a \CW\ laser at 755.6~nm, tuned close to the resonance energy of the bottom layer \RP. The corresponding energy is indicated by a gray vertical line in Fig.~\ref{fig:2}(a). Figure~\ref{fig:2}(b) shows the \RF\ signal at this wavelength as a function of $V_E$. The layer transition takes place between $V_E = 1.9$~V and $V_E=2.3$~V. We measure \RF\ for different electric fields $V_E$ between 1.8~V and 2.4~V. The probe laser power before the cryostat window 
was approximately 20~\si{\micro\watt}. We adjusted the power for each measurement such that the average count rate per \APD\ was always approximately $10^6$~cts/s, meaning that the measurement power increases by approximately a factor of four across the layer transition. The total integration time for each measurement was one hour.

Figure~\ref{fig:2}(c) shows $g^{(2)}(\tau)$ measured at different values of $V_E$. We see a clear bunching signal that is maximal for $V_E=2.1$~V, at the center of the layer transition. We fit each $g^{(2)}(t)$ with the function
\begin{equation}
	f=1+Ae^{-|t-t_0|/t_\text{d}},
\end{equation}
where $A$ is the bunching amplitude, $t_0$ is the time difference originating from different path lengths to the two detectors, and $t_\text{d}$ is the characteristic decay time of the bunching signal. The actual time delay $\tau$ is given by $t-t_0$. The bunching amplitudes obtained from the fits are shown in Fig.~\ref{fig:2}(d). The highest bunching amplitude we measured is $g^{(2)}(0)|_{V_E=2.1\text{ V}}=1.012\pm0.001$.
The maximum in bunching amplitude observed for $V_E = 2.1$~V, where the energies of the electronic states in the two layers are identical, and consequently the layer polarization transition should take place, demonstrates that photon correlation measurements are sensitive to critical fluctuations.

\begin{figure}[h]
    \includegraphics[width=\columnwidth]{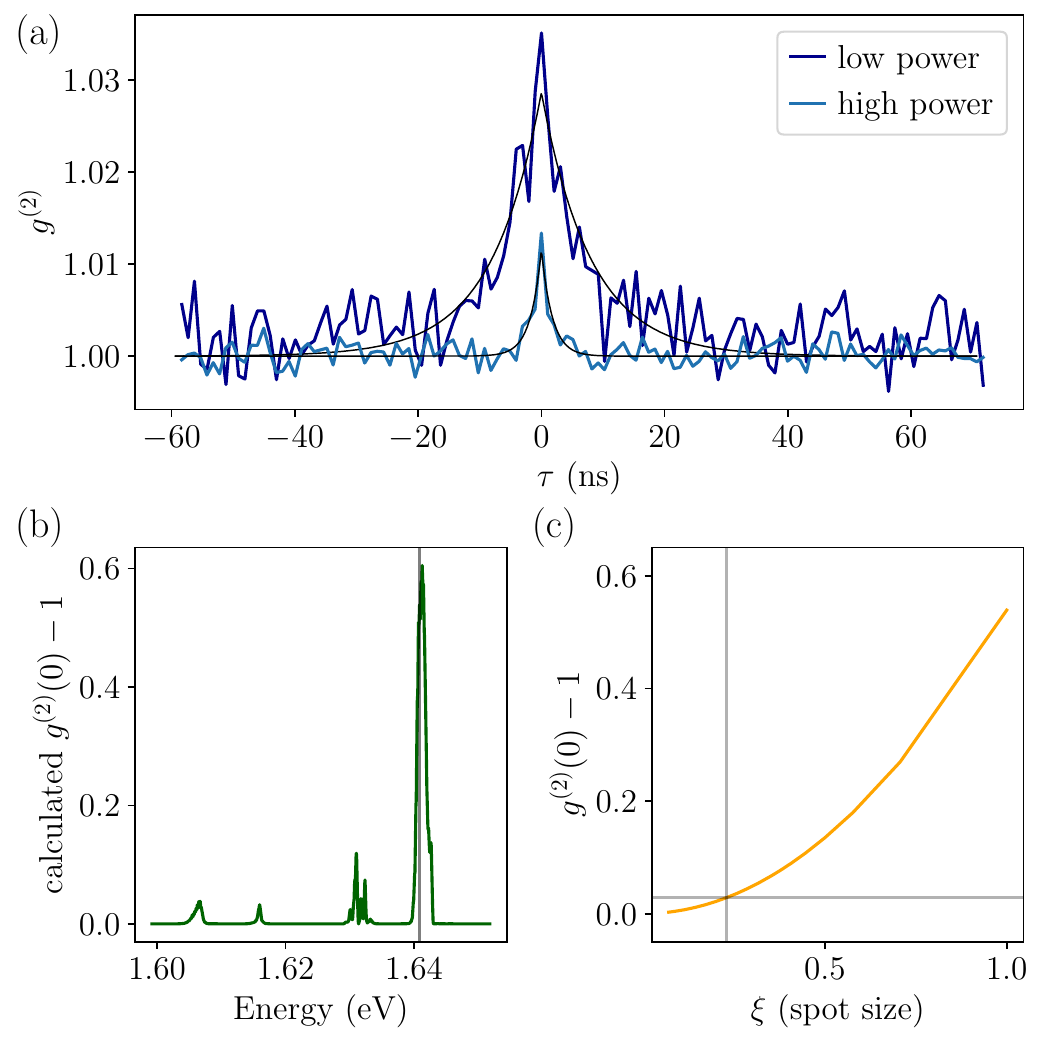}
    \caption{(a) Second-order correlaiton function $g^{(2)}(\tau)$ at the center of the phase transition ($V_E=2.1$~V) for two different excitation powers. The high-power data is the same as shown in Fig. 2(c). The increase in bunching amplitude and decay time at low power indicates that the system is heated by the excitation light in the high-power experiment. (b) Calculated maximal bunching amplitude as a function of excitation light energy assuming $N=1$ (i.e., a single domain within the optical spot), based on the spectra shown in Fig. 2(a) and Eq.~(\ref{eq:g2_rArB}). The energy used in the experiment is indicated by a vertical gray line. (c) Calculated bunching amplitude as a function of correlation length at the measurement light energy. The bunching amplitude $g^{(2)}(0)-1=0.029\pm0.002$ extracted from a fit to the low-power data shown in panel a is indicated by a horizontal gray line. According to this simplified model, it corresponds to a correlation length of approximately 0.23 spot sizes (shown by the vertical gray line). 
    }
    \label{fig:3}
\end{figure}

To assess the role of laser heating, we repeated the measurement at $V_E=2.1$~V with 10 times less power and 10 times longer integration time and observed an increase in bunching amplitude by a factor of 2.4, and in characteristic decay time by a factor of 4.5. Both $g^{(2)}(\tau)$ curves are shown in Fig.~\ref{fig:3}(a). These measurements confirm that the laser absorption leads to an effective increase of the electronic temperature.  Consequently, we conclude that the low power data are taken closer to the phase transition, and the increase in bunching amplitude is a manifestation of the larger correlation length, while the prolonged decay time suggests critical slowing down of the fluctuations. On the other hand, the data is well fitted by a simple exponential function, suggesting that even for the lowest laser power the effective electronic temperature is not close to the Curie temperature. We note that the dependence of the bunching amplitude on the electric field (Fig.~\ref{fig:2}(d)) is asymmetric with respect to the center of the transition, with higher bunching amplitudes observed on the bottom-layer-polarized (higher $V_E$) side. This effect is not a direct consequence of laser heating, as the excitation power was higher on this side of the transition to ensure constant coincidence rates. The reason for the asymmetry remains unclear, but could be related to other light-induced effects such as photo-doping or light-induced interlayer transfer of electrons or holes.
Measuring in the \RF \ configuration is advantageous for correlation measurements due to the large contrast between the layer polarized states and the suppressed background signal (see Fig.~\ref{fig:2}(a)). However, for \RF\ measurements without an external magnetic field the average signal is low, forcing us to use high laser power, which in turn leads to heating. Measuring \RF\ at magnetic fields where the exciton Zeeman splitting exceeds the linewidth of the \RP\ resonance should allow us to substantially lower the incident laser power without compromising the number of coincidences. 

\subsection{Analysis and Modelling}

In the following, we present a simple theoretical model that serves to improve the interpretation of photon correlation measurements. Motivated by the aforementioned experiments, we focus on the critical fluctuations of a system on the verge of a continuous phase transition. We assume the breaking of a discrete $Z_2$ symmetry. For simplicity, we also assume that we have $N$ equally sized domains within the measurement spot, which fluctuate independently between two states $A$ and $B$ with different brightness, as is the case for electronic states polarized in the top and the bottom layer. Each domain has equal probability of being in either state. The system can then be in one of $N+1$ different macroscopic states. We denote the macroscopic state with $k$ domains in the $A$ state and $N-k$ domains in the $B$ state by the index $k$. The degeneracy of state $k$ is $\binom{N}{k}$, while the total number of microscopic states is $2^N$. The occupation probability of the macroscopic state $k$ is therefore $w_k=\binom{N}{k}/2^N$. We denote the intensity of the light scattered by the system as $I(t)=I_\text{in}\cdot r(t)$, where $r(t)=r_k$ with $k\in \{ 0,\ldots, N\}$ quantifies how much of the excitation light intensity $I_\text{in}$ is scattered by the system in state $k$. We assume ergodicity, justifying the replacement of the time by the ensemble average. The bunching amplitude $g^{(2)}(0)$ is then given by:
\begin{equation}
	g^{(2)}(0)=\frac{\langle I(t)^2 \rangle}{\langle I(t)\rangle^2} = \frac{\sum_k w_k(I_\text{in}r_k)^2}{(\sum_k w_k I_\text{in}r_k)^2}=\frac{\sum_k w_k r_k^2}{(\sum_k w_k r_k)^2}.
	\label{eq:g2}
\end{equation}
We assume $r_k=kr_A+(N-k)r_B$, were $r_{A,B}$ is the light scattered by a single domain in state $A,B$. Using $\sum kw_k=N/2$ and $\sum k^2w_k=N(N+1)/4$, we obtain  
\begin{equation}
	g^{(2)}(0)=1+\frac{1}{N}\left(\frac{r_A-r_B}{r_A+r_B}\right)^2.
	\label{eq:g2_rArB}
\end{equation}
Assuming a spot size of 1~\si{\micro\meter}, the correlation length in the system is given by $\xi = \sqrt{1/N}$~\si{\micro\meter}. The closer the system is to the phase transition, the larger the domains (and correlation length), and therefore the lower the number of domains within the spot. Consequently, the bunching amplitude increases as $g^{(2)}(0)\propto\xi^2$.

We use Eq.~(\ref{eq:g2_rArB}) to calculate the maximal possible bunching amplitude in our experiments as a function of the probe laser frequency/photon energy. To this end, we set $N=1$ and use the measured spectra shown in Fig.~\ref{fig:2}(a) as $I_\text{in}r_A$ and $I_\text{in}r_B$. Figure~\ref{fig:3}(b) shows the calculation result: The largest bunching is expected when the probe laser is resonant with the bottom layer \RP. Here, \RF\ is bright when the top layer is doped and almost disappears when the bottom layer is doped. 

We also use Eq.~(\ref{eq:g2_rArB}) to calculate the bunching amplitude as a function of correlation length at the measurement energy. The result is shown in Fig.~\ref{fig:3}(c). The bunching amplitude at low power extracted from a fit to the data in Fig.~\ref{fig:3}(a) is $A_\text{lp}=0.039\pm0.002$. Comparing this value with the theoretical bunching amplitude as a function of correlation length shown in Fig.~\ref{fig:3}(d), we extract a correlation length of 0.23 $\times$ spot size. The measured bunching amplitude and the corresponding correlation length are indicated by gray lines in Fig.~\ref{fig:3}(d).

\section{Conclusions and Outlook}

Our measurements demonstrate that photon correlation spectroscopy can be used to measure electron number fluctuations in the vicinity of a phase transition to a layer-polarized Mott insulator state in a bilayer \TMD\ moiré system. We successfully detected photon bunching in cross-polarized resonance fluorescence detection, and the $V_E$-dependent measurements show that the photon bunching originates from enhancement of fluctuations close to the phase transition. By reducing the laser power, and the associated induced heating or light induced fluctuations of layer pseudo-spin, we approached the critical point. We observed an enhanced bunching signal and a critical slowing down consistent with an enhancement of the correlation length $\xi$ up to $\sim 100$~nanometers, indicating mesoscopic domains extending to tens of moiré sites. We are thus still in a disordered phase, which is consistent with the exponential (instead of algebraic) decay of time correlations, and the lack of hysteresis in the static response to an electric field. Nevertheless, a systematic measurement of the correlation length $\xi$ and time $\tau$ in the disordered phase, as a function of the distance to the phase transition $\delta=T/T_c-1$, could provide a direct measurement of the critical exponents $\nu$ and $z$ defined as $\xi\sim\delta^{-\nu}$ and $\tau\sim\xi^{z}$.

Our experiments highlight the opportunities and challenges of photon correlation spectroscopy for studying phase transitions in strongly correlated electron systems. To measure the photon correlation rate depicted in Figs.~\ref{fig:2} and \ref{fig:3}, we had to apply relatively strong laser intensities $\ge 1$~\si{\micro\watt/\micro\meter^2} to achieve a photon detection rate of $\sim 10^6$~photons/s. We emphasize, however, that achieving such photon count rates does not necessarily require such strong incident laser intensities. If the van der Waals heterostructure is designed to minimize the background reflection from dielectric interfaces and thereby ensure that the detected photons predominantly originate from the exciton resonance, it is possible to obtain count rates $\ge 3 \times 10^6$ photons/s using an incident laser power of only $1$~\si{\pico\watt/\micro\meter^2}. This would correspond to a reduction of more than six orders of magnitude in the incident laser intensity to achieve similar coincidence rates, indicating that laser-induced heating effects could be essentially eliminated. Reducing the requisite laser intensity  while achieving the same coincidence rate in the \RF\ configuration that we have used in our experiments would in turn require an external magnetic field $\ge 5$~T to split the two circularly polarized excitonic resonances by more than their linewidth.

An immediate application of our technique would be in the characterization of quantum phase transitions between different correlated phases in twisted MoTe$_2$ homobilayers. While electronic fluctuations in classical phase transitions originate from thermal fluctuations, an interesting open question is the role of quantum fluctuations that stem from measurement back-action. 

The data that supports the findings of this paper is available in the ETH Research Collection \cite{ethResearchCollection}.

{\sl Acknowledgements} We thank Arthur Christianen, Haydn Adlong and Qi Hu for insightful discussions. This work was supported by the Swiss National Science Foundation (SNSF) under Grant No. 2000-1-240035. K.W. and T.T. acknowledge support from the JSPS KAKENHI (grant numbers 21H05233 and 23H02052), the CREST (JPMJCR24A5), JST and World Premier International Research Center Initiative (WPI), MEXT, Japan.

\appendix

\section{Experimental Setup and Methods}

The experiments were performed in a confocal microscope setup, with the sample cooled to $T=4.2$~K in a liquid He bath cryostat. We used a single-mode-fiber-coupled broadband \LED\ to measure broadband spectra. For photon correlation measurements, we used a tunable \CW\ Ti:Sapphire laser. The light emitted by the laser is coherent, so the excitation has completely uncorrelated photons. Any deviation of $g^{(2)}$ from 1 is therefore caused by interaction of the light with the sample. The light was focused on the sample using a long-working-distance apochromatic cryogenic objective. Spectra were measured using a 0.75~m spectrometer equipped with a liquid-N$_2$-cooled \CCD\ camera and a 1200~groove/mm diffraction grating. For photon correlation measurements, we used a Hanbury-Brown--Twiss setup with two \APD s and a time-correlated single photon counting system. To calculate the second-order correlation function $g^{(2)}(\tau)$, we only use two-detector coincidences, as the measured correlation times are smaller than the \APD\ dead time of approximately 85~ns. We calculate $g^{(2)}$ using a binning time of 1.024~ns and a maximal time delay of approximately 72~ns.

\section{Signal and integration time in selected experimentally relevant scenarios}
\label{sec:app_theory}

We now more generally discuss concrete experimental scenarios inspired by typical \TMD\ spectral features to highlight some important considerations when designing samples to be studied using photon correlation spectroscopy. We limit our discussion to the case of one domain within the optical spot ($N=1$), and define the signal $S$ as the bunching amplitude at zero time delay (see Eq.~(\ref{eq:g2_rArB})):
\begin{equation}
    S=g^{(2)}(0)-1=\left(\frac{I_A-I_B}{I_A+I_B}\right)^2.
    \label{eq:S_I}
\end{equation}
As discussed in the main text, the scattered light intensity in state $i$ is given by $I_i=r_i\cdot I_\text{in}$, with $I_\text{in}$ the excitation light intensity. Since the parameters $r_i$ are constants given by the device properties, the only source of uncertainty on $S$ stems from the shot noise of the incoming light field $I_\text{in}$. We therefore rewrite Eq.~(\ref{eq:S_I}) in terms of photon numbers $N_i$:
\begin{equation}
    S=\left(\frac{N_A-N_B}{N_A+N_B}\right)^2,
    \label{eq:S_N}
\end{equation}
where we have used $I_i\propto N_i$. The noise level is therefore given by
\begin{align}
    \Delta S
    &=\sqrt{\left(\frac{\partial S}{\partial N_A}\right)^2\Delta N_A^2+\left(\frac{\partial S}{\partial N_B}\right)^2\Delta N_B^2}\\
    &=\frac{4\sqrt{2}r_Ar_B|r_A-r_B|}{(r_A+r_B)^3}\cdot\frac{1}{\sqrt{N_\text{in}}},
\end{align}
where we have used $N_i=r_iN_\text{in}$ and $\Delta N_\text{in}=\sqrt{N_\text{in}}$. To reach a relative uncertainty $\Delta S/S=\kappa$, we therefore need to send
\begin{align}
    N_\text{in}=\frac{32r_A^2r_B^2}{(r_A^2-r_B^2)^2\kappa^2}
    \label{eq:Nin}
\end{align}
photons into the system. The input photon number $N_\text{in}$ is proportional to $I_\text{in}\cdot T$, where $T$ is the integration time.

Resonances in \DR\ spectra of \TMD s can to first order be described by a dispersive Lorentzian function
\begin{align}
	\frac{r-r_0}{r_0}=&L_\text{disp}(E;E_0,\gamma,A,\alpha)\\
    =&A\cos(\alpha)\frac{\gamma/2}{(E-E_0)^2+(\gamma/2)^2}\notag\\
	&+A\sin(\alpha)\frac{E_0-E}{(E-E_0)^2+(\gamma/2)^2},
	\label{eq:disp_lorentzian}
\end{align}
where $E_0, \gamma$, and $A$ are the resonance energy, linewidth, and amplitude. The phase $\alpha$ is determined by the interference of the light that is scattered by the \TMD\ with reflections from other surfaces in the heterostructure. Depending on the value of the $\alpha$, the resonance shape can range from a perfectly non-dispersive Lorentzian peak ($\alpha=0$) to a dip ($\alpha = \pi$), and everything in between. It is usually desirable to have non-dispersive line shapes, meaning either perfect peaks or dips. A desired line shape can be achieved by choosing an appropriate substrate and encapsulating \hBN\ flakes of suitable thickness. By tailoring the distance between interfaces in the device, it is possible to set the phase difference between the reflected signals from each interface. To achieve a peak line shape, \hBN\ thicknesses should be chosen such that the background reflectivity is minimized. For a dip line shape, the background reflectivity is maximal.

\begin{figure}[h]
    \centering
    \includegraphics[width=\columnwidth]{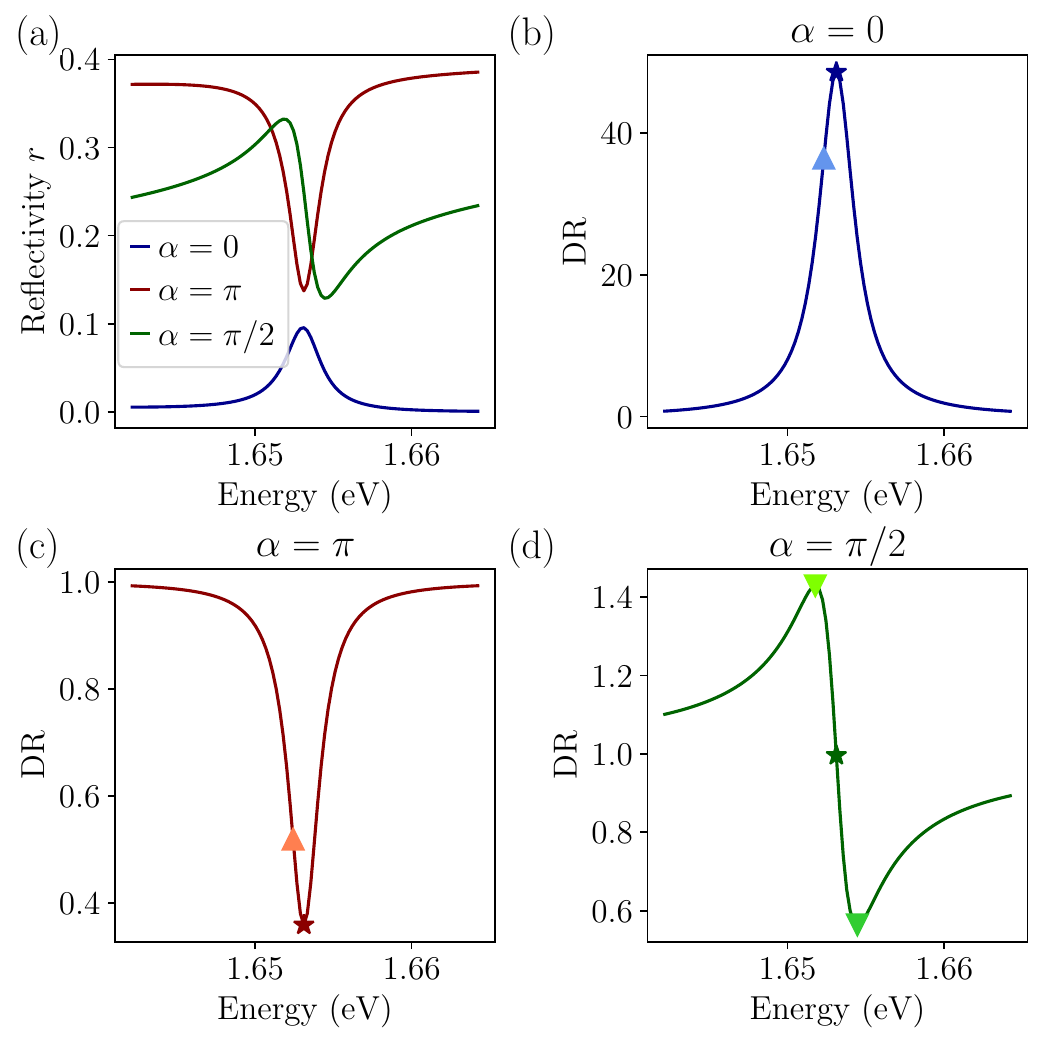}
    \caption{MoSe$_2$ reflectance spectra obtained using transfer matrix simulations. We chose encapsulating h-BN thicknesses to obtain a peak ($\alpha=0$), a dip ($\alpha=\pi$), and a purely dispersive ($\alpha=\pi/2$) line shape. (a) Reflectivity $r$ as a function of photon energy in the three cases, calculated using the transfer matrix formalism. (b-d) Differential reflectance obtaine by fitting the normalized transfer matrix result, $(r-r_0)/r_0$, with dispersive Lorentzian functions. The DR signal on resonance is marked by a star. In panels (b) and (c), the photon energy and DR signal on the low-energy flank of the resonance, where the slope is maximal, is indicated by a triangle. In panel (d), the reflectivity minimum and maximum are marked by triangles.}
    \label{fig:SI_specs}
\end{figure}

To compare different line shapes in the context of photon correlation measurements, we use the transfer matrix method to calculate three realistic reflectance spectra: a peak, a dip, and a maximally dispersive line shape. We simulate a van der Waals heterostructure consisting of a \mose\ layer encapsulated by top and bottom \hBN\ flakes with thicknesses $d_\text{t}$ and $d_\text{b}$, respectively, and top and bottom monolayer graphene gates. The heterostructure is placed on top of a Si substrate capped by a 285~nm-thick SiO$_2$ layer. We find a peak line shape ($\alpha=0$) for $d_\text{t}=30$~nm, $d_\text{b}=27$~nm, a dip ($\alpha=\pi$) for $d_\text{t}=87$~nm, $d_\text{b}=56$~nm, and a purely dispersive line shape ($\alpha=\pi/2$) for $d_\text{t}=16$~nm, $d_\text{b}=87$~nm. The reflectivities for the three structures are shown in Fig.~\ref{fig:SI_specs}(a). We further calculate the background reflectivity $r_0$ using the same transfer matrix method, by removing the \TMD\ layer from the heterostructure. This allows us to calculate the \DR\ spectra for each structure, which we show in Fig~\ref{fig:SI_specs}(b-d) and fit with the dispersive Lorentzian function $L_\text{disp}$ (see Eq.~(\ref{eq:disp_lorentzian})). We now use Eqs.~(\ref{eq:S_I}) and (\ref{eq:Nin}) to calculate the bunching amplitude $S$ and the input photon number $N_\text{in}$ (as a proxy for integration time) for different photon energies and fluctuation types. For the nondispersive line shapes, we consider the cases of measuring exactly on resonance ($E=E_0$) and on the low-energy flank of the resonance ($E=E_0-\sqrt{3}\gamma/6$), where the slope of the Lorentzian is maximal. These energies are marked by stars and triangles, respectively, in Figs.~\ref{fig:SI_specs}(b) and (d). For the dispersive line, we also consider the measurement on resonance ($E=E_0$) and at the minimum and maximum of reflectivity ($E=E_0\pm\gamma/2$). The energies are marked by a star and triangles in Fig.~\ref{fig:SI_specs}(d).

\begin{figure}[h]
    \centering
    \includegraphics[width=\columnwidth]{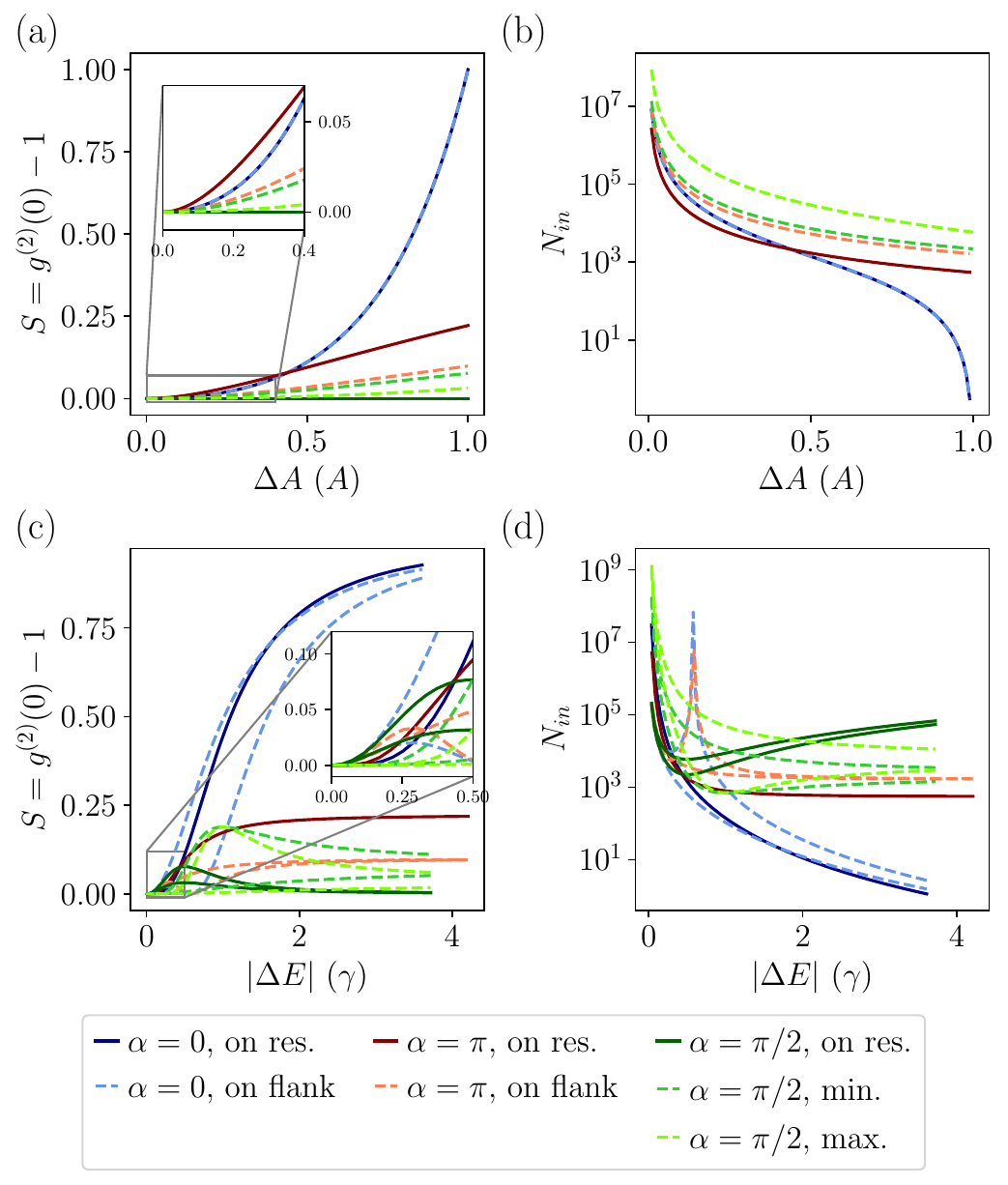}
    \caption{Calculated bunching amplitudes $S$ and input photon numbers $I_\text{in}$ necessary to achieve a relative uncertainty of $\kappa=0.1$ for different resonance line shapes and photon energies. (a-b) Bunching amplitude $S$ and input photon number $N_\text{in}$ as a function of resonance amplitude fluctuations $\Delta A$. (c-d) Bunching amplitude $S$ and input photon number $N_\text{in}$ as a function of resonance energy fluctuations $\Delta E$. The insets in panels (a) and (c) show details in the small fluctuation limit.}
    \label{fig:SI_calcg2}
\end{figure}

We separately study the two cases of amplitude fluctuations $\Delta A$ and energy fluctuations $\Delta E$. Figures~\ref{fig:SI_calcg2}(a) and (b) show $S$ and $N_\text{in}$ as a function of $\Delta A$. For the calculation of $N_\text{in}$, we chose $\kappa=0.1$. We see that for the on-resonance measurement of the dispersive resonance, $S=0$ $\forall\Delta A$, because the reflectivity at the resonance energy is always equal to $r_0$. For large $\Delta A\approx A$, measuring either on resonance or on the flank with a peak line shape gives the largest $S$ and smallest $N_\text{in}$. However, for small $\Delta A\lesssim 0.5 A$, measuring on resonance with a dip line shape is preferable. Figures~\ref{fig:SI_calcg2}(c) and (d) show $S$ and $N_\text{in}$ as a function of $|\Delta E|$. When the resonance is asymmetric with respect to the measurement point, results differ for positive and negative $\Delta E$, and the dependence of $S$ and $N_\text{in}$ on $\Delta E$ is in general not monotonous. We see that for large fluctuations $|\Delta E|\gg\gamma$, the peak line shape is again the most advantageous. For small fluctuations, measuring on the flank of the peak yields the highest $S$ and smallest $N_\text{in}$, while the dispersive line on resonance and the dip on resonance or on the flank also beat the peak on resonance. Bunching amplitudes from dispersive lines have a maximum at $|\Delta E|=\gamma/2$ or $|\Delta E|=\gamma$ for measurements on resonance or at the extrema of reflectivity, respectively, as for these $|\Delta E|$ values the maximal contrast $|I_A-I_B|$ is achieved. 

We conclude that when fluctuations are expected to be large, i.e., leading to spectral fluctuations on the order of the resonance oscillator strength or linewidth, the best results are achieved at the resonance energy for nondispersive peaks. However, when the fluctuations are small, other line shapes and measurement energies can be advantageous.

We further note that if the resonance has well-defined polarization selection rules, it is also possible to achieve a non-dispersive Lorentzian peak line shape by exciting and detecting with two orthogonal polarizations that each have a polarization vector component along the polarization axis of the resonance, as we demonstrated in the experiments presented in the main text. For example, resonances in \TMD s in sufficiently large out-of-plane magnetic fields are circularly polarized due to spin-valley locking. In linear cross-polarized excitation and detection, the background signal is filtered out while resonances appear as Lorentzian peaks.  
The main difference between cross- and co-polarized detection is that a higher background suppression is typically achievable in cross-polarized detection. Transfer matrix simulations show that the reflectivity at $\lambda=755$~nm (the typical resonance wavelength of the \mose\ A exciton) from a \hBN\ flake on top of a routinely used 285~nm SiO$_2$-capped Si wafer reaches a minimum of approximately $10^{-3}$ for an \hBN\ thickness of 61~nm. If graphene top and bottom gates are added, the background reflectivity increases. In contrast, a polarization suppression on the order of  $10^{-4}-10^{-5}$ is usually easily achievable at a single frequency. A smaller background reflectivity $r_0$ is equivalent to a larger amplitude $A$ of the dispersive Lorentzian. Therefore, even though the maximal achievable signal in cross-polarized detection is only $1/4$ of the one in co-polarized detection, the reduction in background by at least an order of magnitude implies that larger bunching amplitudes are achievable in cross-polarized detection.

The above discussion is highly simplified compared to a real experiment. For example, fluctuations usually do not affect only one parameter of the resonance, and a measured intensity is never exactly zero. Nevertheless, it can serve to illustrate some of the less intuitive results one can expect from photon correlation spectroscopy experiments on \TMD s. Equation~(\ref{eq:g2_rArB}) can be used to calculate the expected bunching amplitude based on measurement spectra when these are available on both sides of the phase transition that is studied, as demonstrated in the main text.



%

\end{document}